\documentclass{aa}
\usepackage{natbib}
\usepackage{txfonts}
\usepackage{hyperref}
\usepackage{graphicx}
\graphicspath{{figures/}}

\newcommand*{\hd}{HD~43317\xspace}

\begin{document}

\title{Period spacings of gravity modes in rapidly rotating magnetic stars}
\subtitle{II.~The case of an oblique dipolar fossil magnetic field}

\author{V. Prat\inst{1} \and S. Mathis\inst{1,2} \and C. Neiner\inst{2} \and J. Van Beeck\inst{3} \and D. M. Bowman\inst{3} \and C. Aerts\inst{3,4,5}}

\institute{
    AIM, CEA, CNRS, Universit\'e Paris-Saclay, Universit\'e Paris Diderot, Sorbonne Paris Cit\'e, F-91191 Gif-sur-Yvette, France
    \and
    LESIA, Paris Observatory, PSL University, CNRS, Sorbonne Universit\'e, Universit\'e de Paris, 5 place Jules Janssen, 92195 Meudon, France
    \and
    Instituut voor Sterrenkunde, KU Leuven, Celestijnenlaan 200D, 3001 Leuven, Belgium
    \and
    Dept. of Astrophysics, IMAPP, Radboud University Nijmegen, 6500 GL, Nijmegen, The Netherlands
    \and
    Max Planck Institute for Astronomy, Koenigstuhl 17, 69117 Heidelberg, Germany
}

\date{}

\abstract
{Stellar internal magnetic fields have recently been shown to leave a detectable signature on period spacing patterns of gravity modes.}
{We investigate the effect of the obliquity of a mixed (poloidal and toroidal) dipolar internal fossil magnetic field with respect to the rotation axis on the frequency of gravity modes in rapidly rotating stars.}
{We use the traditional approximation of rotation to compute non-magnetic modes, and a perturbative treatment of the magnetic field to compute the corresponding frequency shifts.
We apply the new formalism to \hd, a magnetic, rapidly rotating, slowly pulsating B-type star, whose field has an obliquity angle of about $80^\circ$.}
{We find that frequency shifts induced by the magnetic field on high-radial-order gravity modes are larger with increasing obliquity angle, when the magnetic axis is closer to the equatorial region, where these modes are trapped.
The maximum value is reached for an obliquity angle of $90^\circ$.
This trend is observed for all mode geometries.}
{Our results predict that the signature of an internal oblique dipolar magnetic field is detectable using asteroseismology of gravity modes.}


\maketitle

\section{Introduction}

Thanks to high-precision space photometry with missions such as CoRoT \citep{Baglin}, \emph{Kepler} \citep{Borucki}, K2 \citep{Howell}, and now TESS \citep{Ricker}, the study of stellar pulsations has rapidly developed over the last 15 years.
It has provided important new information on the internal structure of stars through seismic modelling \citep[see][for a recent review]{Aerts19}.
In parallel, the observational study of magnetism in stars, e.g. through spectropolarimetry from ground-based facilities, has provided major results on surface magnetic fields and rotation \citep[e.g.][]{DonatiLandstreet, Neiner15, Wade}.
Studying both physical processes (pulsations and magnetism) at the same time offers further opportunities.
However, this new combined technique, called magneto-asteroseismology, requires the identification of diagnostic tools.

As a first step, in \cite{paper1} (hereafter Paper~I), we investigated the effect of an axisymmetric mixed (with both poloidal and toroidal components) magnetic field on gravito-inertial modes computed in the traditional approximation of rotation \citep[TAR; e.g.][]{LeeSaio97, Townsend03, Bouabid}.
Theoretical and numerical studies have shown that the stability of fossil magnetic fields requires such a mixed configuration \citep[e.g.][]{Tayler80, BraithwaiteSpruit, Braithwaite08, DuezBM}.
We found that a dipolar field with a near-core strength of order 100\,kG, corresponding to a few kG at the surface as typically observed in approximately 10\% of hot stars \citep{shultz2019}, shifts the frequencies of gravity modes and modifies the period spacing patterns, which allow us to probe near-core mixing and rotation \citep{VanReeth15a, VanReeth15b, Ouazzani17, Christophe}, in a way that should be detectable in current observations.

However, spectropolarimetric surveys show that most stars with a fossil magnetic field have a tilted magnetic axis with respect to the rotation axis \citep[e.g.][]{grunhut2015}.
This obliquity may affect the frequency shifts induced by the magnetic field.
Therefore, in the present paper, we address the effect of an oblique mixed magnetic field on the frequency of gravito-inertial modes.

\section{Frequency shifts}
\label{sec:shifts}

In the present work we consider a magnetic field that is weak enough such that the effect of the unperturbed Lorentz force on the hydrostatic equilibrium state is negligible \citep[e.g.][]{DuezMTC}.
We use the perturbation theory for rotating stars presented in Paper~I to compute the frequency shifts induced by the perturbed Lorentz force
\begin{equation}
    \delta\vec F_{\rm L} = \frac{1}{\mu_0}[(\vec\nabla\wedge\vec B)\wedge\delta\vec B + (\vec\nabla\wedge\delta\vec B)\wedge\vec B],
\end{equation}
where $\mu_0$ is the vacuum permeability, $\vec B$ is the large-scale magnetic field, and $\delta\vec B$ are the fluctuations of the magnetic field due to the oscillation displacement.
Those are given in the adiabatic case by the induction equation
\begin{equation}
    \delta\vec B = \vec\nabla\wedge(\vec\xi_0 \wedge \vec B),
\end{equation}
where $\vec\xi_0$ is the unperturbed displacement of the mode.
The frequency shifts induced by the magnetic field read
\begin{equation}
    \label{eq:gen_split}
    \delta\omega = -\frac{\langle\vec\xi_0,\delta\vec F_{\rm L}/\rho\rangle}{2\omega_0\langle\vec\xi_0,\vec\xi_0\rangle+\langle\vec\xi_0,2i\vec\Omega\wedge\vec\xi_0\rangle},
\end{equation}
where $\rho$ is the density of the background model, $\omega_0$ is the unperturbed angular frequency of the mode in the corotating frame, $\vec\Omega$ is the rotation vector, the scalar product is defined by
\begin{equation}
    \langle\vec\xi,\vec\zeta\rangle=\int_V\rho\vec\xi^*\cdot\vec\zeta{\rm d}V,
\end{equation}
and the asterisk ($^*$) denotes the complex conjugate.

In the TAR, the horizontal component of the rotation vector is neglected ($\vec\Omega\simeq\Omega\cos\theta\vec e_{\rm r}$), and unperturbed eigenmodes for gravito-inertial waves are given in spherical coordinates $(r,\theta,\varphi)$ by
\begin{equation}
    \vec\xi_0 = [\xi_{\rm r}(r) H_{\rm r}(\theta)\vec e_{\rm r} + \xi_{\rm h}(r) H_\theta(\theta)\vec e_\theta + i\xi_{\rm h}(r) H_\varphi(\theta)\vec e_\varphi]e^{i(m\varphi-\omega_0 t)},
\end{equation}
where $m$ is the azimuthal order, $\vec e_{\rm r}$, $\vec e_\theta$, and $\vec e_\varphi$ are radial, latitudinal, and azimuthal unit vectors, and $H_{\rm r}$, $H_\theta$, and $H_\varphi$ are radial, latitudinal, and azimuthal Hough functions, respectively \citep{Hough,LeeSaio97, Townsend03}.
Their precise definitions are given in Paper~I.

\begin{figure}
    \resizebox{\hsize}{!}{\includegraphics{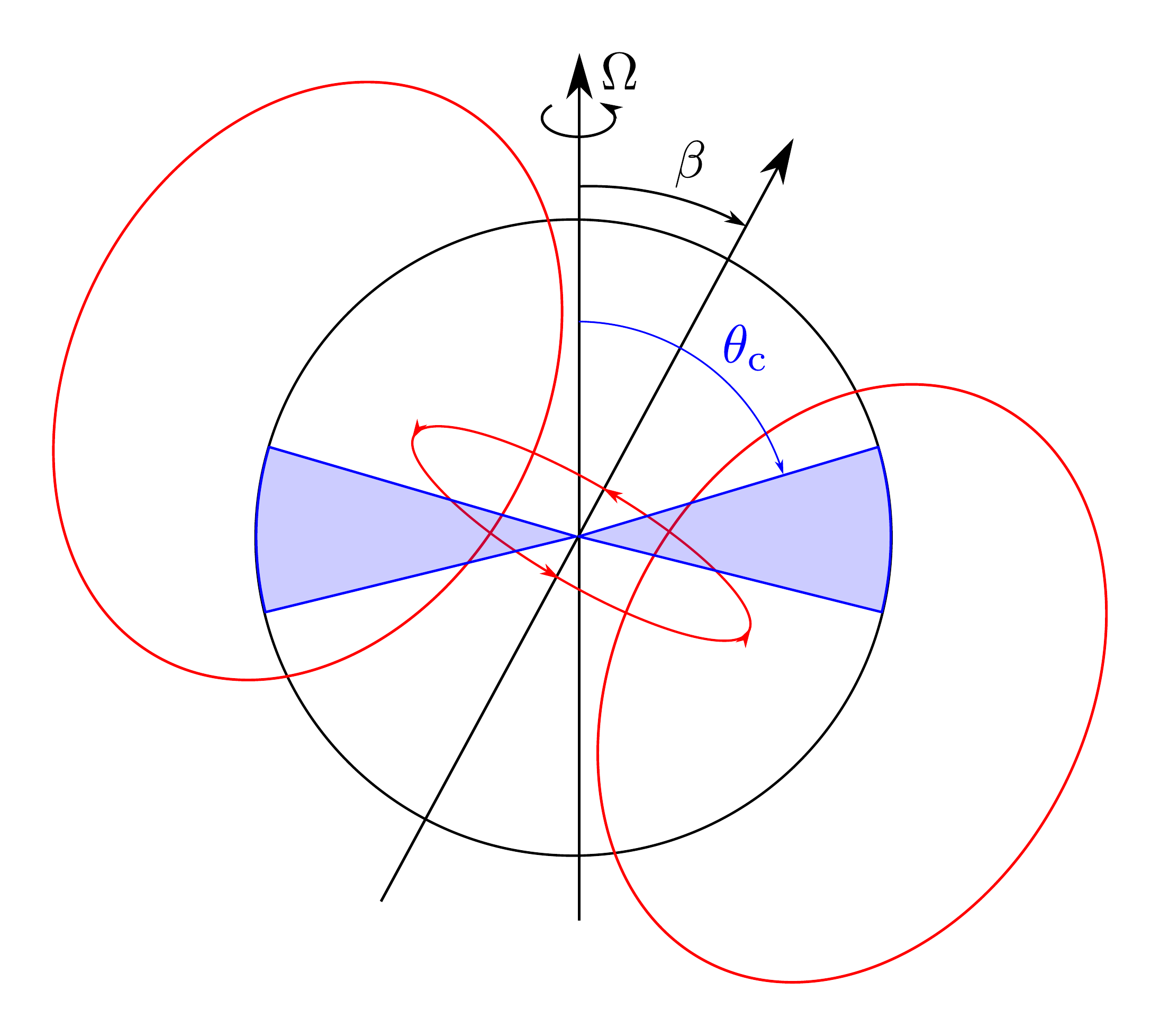}}
    \caption{Sketch of the field geometry. The magnetic field lines (decomposed into poloidal and toroidal components) are drawn in red. An example of a critical latitude for gravito-inertial waves is drawn in blue.}
    \label{fig:schema}
\end{figure}

In the present study we consider the magnetic field as a small pertubation of a rapidly rotating system.
Thus, the oscillation axis can be approximated as the rotation axis.
In the magnetic frame, which is inclined by an angle $\beta$ with respect to the rotation axis as illustrated in Fig.~\ref{fig:schema}, the magnetic field reads
\begin{equation}
    \vec B = B_0[b_{\rm r}(r)\cos\theta'\vec e_{\rm r} + b_\theta(r)\sin\theta'\vec e_{\theta'} + b_\varphi(r)\sin\theta'\vec e_{\varphi'}],
\end{equation}
where $B_0$ is the strength of the magnetic field, $b_{\rm r}$, $b_\theta$, and $b_\varphi$ are radial functions, detailed in Paper~I, that model realistic stable fossil configurations \citep{DuezBM}, and the spherical coordinates in the magnetic frame $(\vec e_{\rm r}, \vec e_{\theta'}, \vec e_{\varphi'})$ are $(r,\theta',\varphi')$.
In the corotating frame, this leads to
\begin{equation}
    \vec B = \vec B_{\rm pol}\cos\beta + \vec B_{\rm eq}\sin\beta,
\end{equation}
where
\begin{equation}
    \vec B_{\rm pol} = B_0(b_{\rm r}\cos\theta\vec e_{\rm r} + b_\theta\sin\theta\vec e_\theta + b_\varphi\sin\theta\vec e_\varphi)
\end{equation}
is axisymmetric (with the axis aligned with the polar axis), and
\begin{equation}
    \begin{aligned}
        \vec B_{\rm eq} &=  B_0[b_{\rm r}\sin\theta\cos\varphi \vec e_{\rm r} - (b_\theta\cos\theta\cos\varphi+b_\varphi\sin\varphi)\vec e_\theta\\
                        &\quad+ (b_\theta\sin\varphi-b_\varphi\cos\theta\cos\varphi)\vec e_\varphi]
    \end{aligned}
\end{equation}
is fully oblique (with the axis in the equatorial plane).

Because of the dependence of $\vec B_{\rm eq}$ on $\varphi$, when computing $\langle\vec\xi_0,\delta\vec F_{\rm L}/\rho\rangle$, the integral over $\varphi$ cancels out mixed terms in $\vec B_{\rm pol}$ and $\vec B_{\rm eq}$, thus leading to
\begin{equation}
    \langle\vec\xi_0,\delta\vec F_{\rm L}/\rho\rangle = \langle\vec\xi_0,\delta\vec F_{\rm L}^{\rm pol}/\rho\rangle\cos^2\beta + \langle\vec\xi_0,\delta\vec F_{\rm L}^{\rm eq}/\rho\rangle\sin^2\beta,
\end{equation}
where $\delta\vec F_{\rm L}^{\rm pol}$ and $\delta\vec F_{\rm L}^{\rm eq}$ are the perturbed Lorentz forces based on $\vec B_{\rm pol}$ and $\vec B_{\rm eq}$ only, respectively.
For high-radial-order modes, the eigenfunctions are rapidly oscillating in the radial direction, and the dominant terms in the full expansion given in Appendix~\ref{sec:terms} yield
\begin{equation}
    \label{eq:freq_ratio}
    \frac{\delta\omega}{\omega_0}=\frac{B_0^2}{2\mu_0\omega_0^2\rho_{\rm c}R^2}I_{\rm r}I_\theta,
\end{equation}
where $R$ is the stellar radius,
\begin{equation}
    \label{eq:Ir}
    I_{\rm r} = \frac{\int_0^1|{\rm d}(xb_{\rm r}\xi_{\rm h})/{\rm d}x|^2{\rm d}x}{\int_0^1|\xi_{\rm h}|^2(\rho/\rho_{\rm c})x^2{\rm d}x},
\end{equation}
with $x=r/R$,
\begin{equation}
    \label{eq:Itheta}
    I_\theta = \frac{\int_0^\pi(H_\theta^2+H_\varphi^2)(\cos^2\theta\cos^2\beta+\frac12\sin^2\theta\sin^2\beta)\sin\theta{\rm d}\theta}{\int_0^\pi(H_\theta^2+H_\varphi^2+\nu H_\theta H_\varphi\cos\theta)\sin\theta{\rm d}\theta},
\end{equation}
and $\nu=2\Omega/\omega_0$ is the spin factor.
Equation~\eqref{eq:freq_ratio} shows that the magnetic frequency shifts scale with the square of the field strength, as in the axisymmetric case.
In addition, Eq.\eqref{eq:Itheta} implies that the dependence on $\beta$ of the frequency shifts induced by the magnetic field is of the form
\begin{equation}
    \label{eq:dep_beta}
    \frac{\delta\omega}{\omega_0} = \left(\frac{\delta\omega}{\omega_0}\right)_{\rm pol}\cos^2\beta + \left(\frac{\delta\omega}{\omega_0}\right)_{\rm eq}\sin^2\beta.
\end{equation}

\section{Application to \hd}
\label{sec:appl}

Similarly to Paper~I, we now compute the new frequency shifts for a representative stellar model of \hd, which is a rapidly rotating, magnetic, slowly pulsating B-type star exhibiting \emph{g} modes \citep{Buysschaert18}.
The adopted model has a mass of $5.8\,{\rm M}_\odot$, a radius of $3.39\,{\rm R}_\odot$, an effective temperature of 17822\,K, a solar-like metallicity, and a central hydrogen mass fraction of 0.54, which corresponds to an age of $28.4\,{\rm Myr}$.
The identified \emph{g}-mode frequencies of \hd range from $0.69162\,{\rm d}^{-1}$ to $5.00466\,{\rm d}^{-1}$.
The rotation period of the star is $P_{\rm rot}=0.897673(4)\,\rm{d}$ \citep{Papics12}, which corresponds to 33\% of the Keplerian angular velocity.
Its mostly dipolar surface magnetic field has a strength of $1312\pm332\,{\rm G}$ and an obliquity angle of $81\pm6^\circ$ \citep{Buysschaert18}.
However, we investigate here how the frequency shifts vary with the obliquity.
Therefore, we consider the latter as a variable of the problem hereafter.
For non-axisymmetric modes, it is important to distinguish between the frequency in the corotating frame $\omega$ and the frequency in the inertial frame $\omega_{\rm i}$, where
\begin{equation}
    \label{eq:doppler}
    \omega_{\rm i} = \omega + m\Omega.
\end{equation}
For non-axisymmetric magnetic fields, the observed frequencies are different from the intrinsic frequencies in the inertial frame, due to the fact that the rotation axis is not exactly the same as the oscillation axis.
However, since we consider only the perturbative effect of the magnetic field while rotation is treated non-perturbatively, we choose to neglect this effect and impose that the oscillation axis is the same as the rotation axis.

In this section, we investigate the effect of an oblique magnetic field on zonal (Sect.~\ref{sec:zonal}), prograde (Sect.~\ref{sec:prog}), and retrograde (Sect.~\ref{sec:retro}) dipole modes.
The impact of rotation is discussed in Sect.~\ref{sec:rot}.

\subsection{Zonal modes}
\label{sec:zonal}

\begin{figure}
    \resizebox{\hsize}{!}{\includegraphics{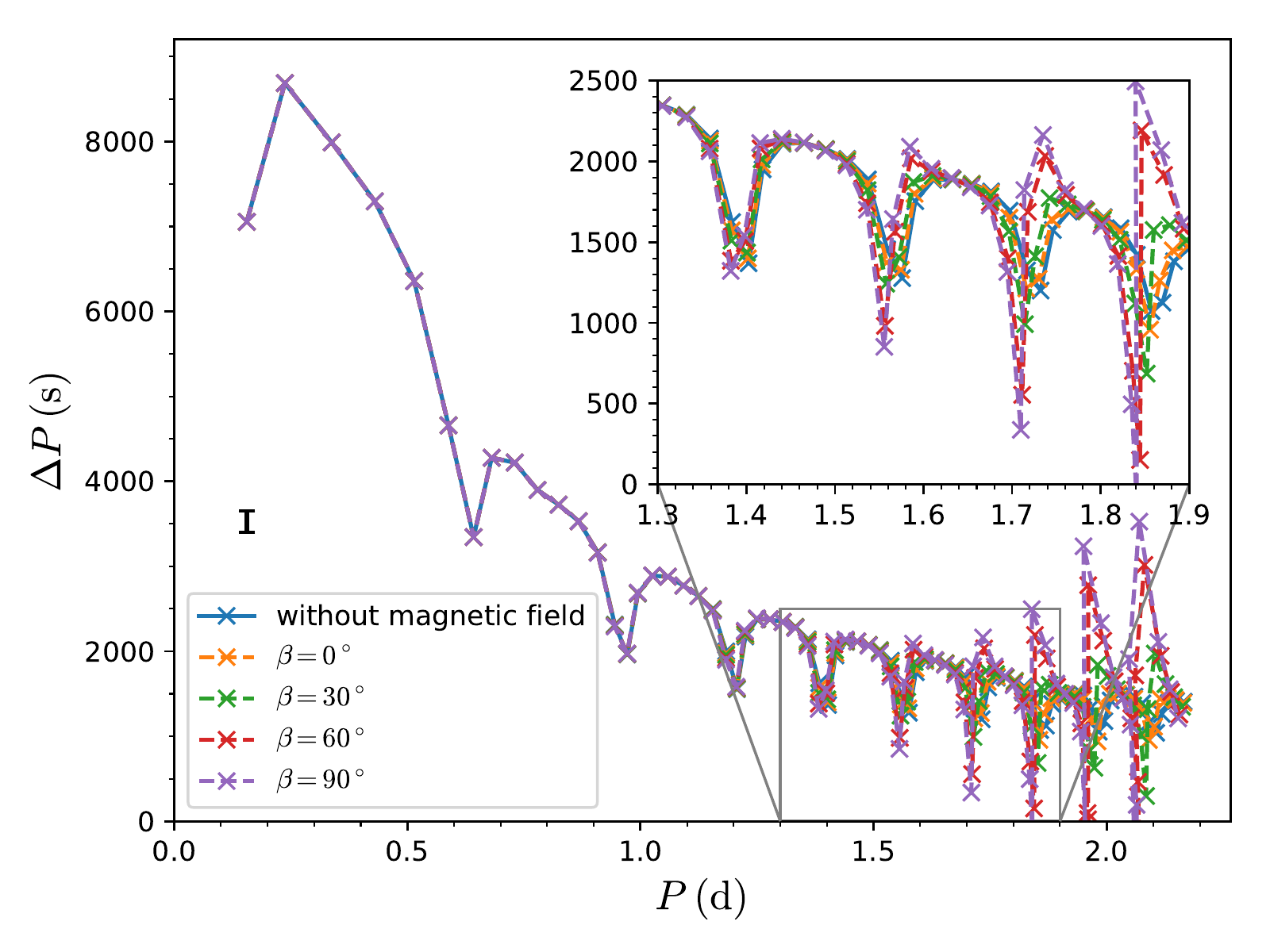}}
    \caption{Period spacings of \emph{g} modes with radial orders from -1 (left) to -74 (right) as a function of the period in the inertial frame for $\ell=1$, $m=0$, $B_0=10^5\,{\rm G}$, and different obliquity angles.
    The vertical black bar on the left represents a typical observational error bar of 250\,s \citep{VanReeth15b}.}
    \label{fig:l1m0}
\end{figure}

\begin{figure}
    \resizebox{\hsize}{!}{\includegraphics{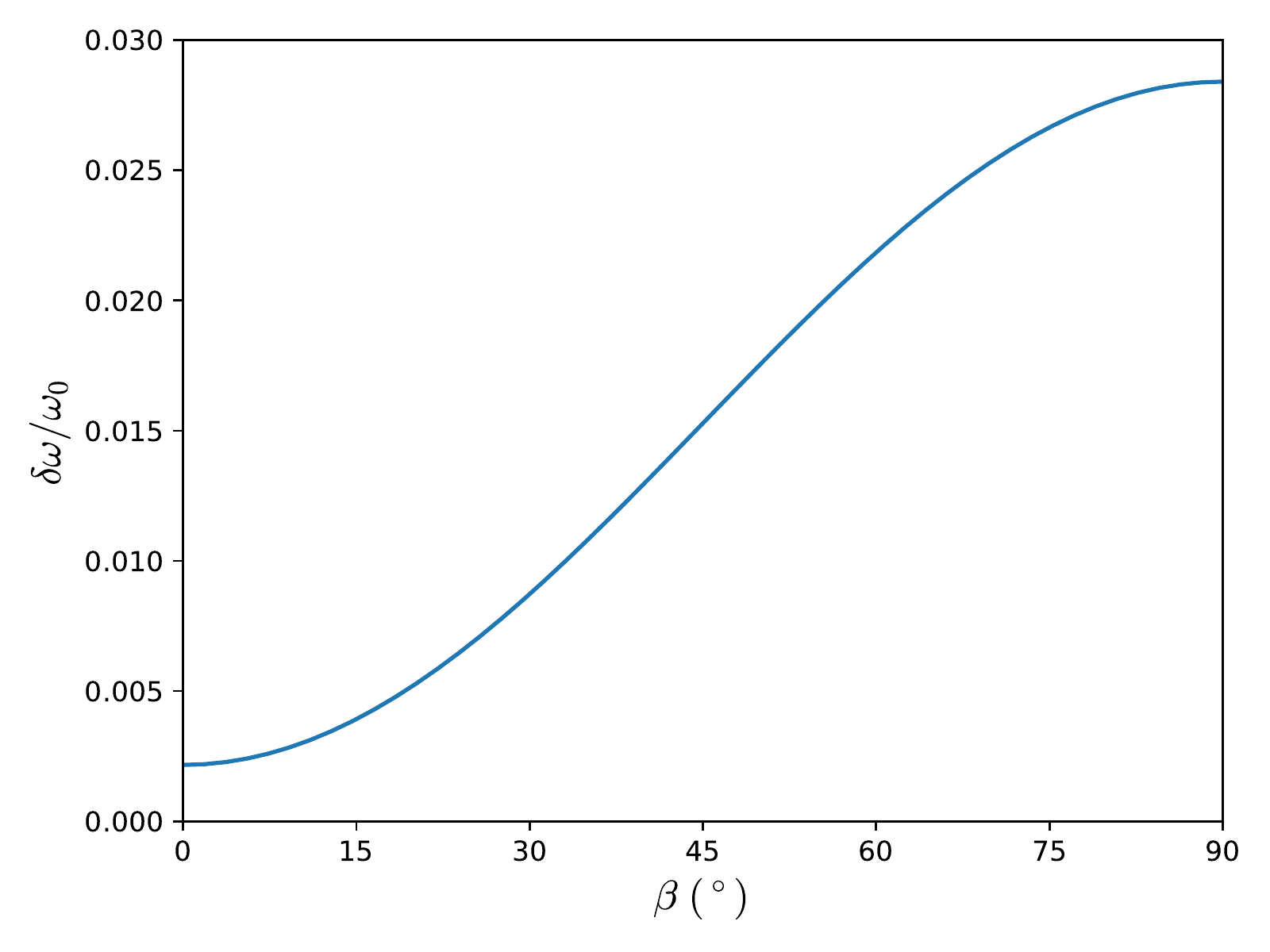}}
    \caption{Relative frequency shift induced by the magnetic field of the mode $\ell=1$, $m=0$, $n=-69$ as a function of the obliquity angle for $B_0=10^5\,{\rm G}$.
    This mode has an unperturbed period of about $2.12\,{\rm d}$.}
    \label{fig:dep_beta}
\end{figure}

Figure~\ref{fig:l1m0} represents for $\ell=1$ and $m=0$ the period spacing $\Delta P$ between modes of consecutive radial orders as a function of the period $P=2\pi/\omega_{\rm i}$.
The first result is that the period spacing patterns for an oblique field are very similar to those obtained in the axisymmetric case.
Indeed, the dips related to near-core chemical gradients are transformed at long periods into a sawtooth-like pattern.
In addition, it shows that oblique fields induce stronger signatures in the period spacing patterns than axisymmetric fields, with a maximum for an obliquity angle of $90^\circ$.
This can be explained by the fact that long-period gravity modes, which are most affected by the magnetic field, are trapped near the equatorial plane, as illustrated in Fig.~\ref{fig:schema}.
These modes are mainly sensitive to the radial component of the magnetic field, as shown by Eq.~\eqref{eq:Ir}, which is maximal on the magnetic axis.
Therefore, the expected signature is stronger when the magnetic axis is closer to the equatorial plane, with a maximum for an obliquity angle of $90^\circ$.
For the $n=-69$ mode, the dependence of the frequency shift induced by the magnetic field on $\beta$, as predicted by Eq.~\eqref{eq:dep_beta}, is shown in Fig.~\ref{fig:dep_beta}.

\begin{figure}
    \resizebox{\hsize}{!}{\includegraphics{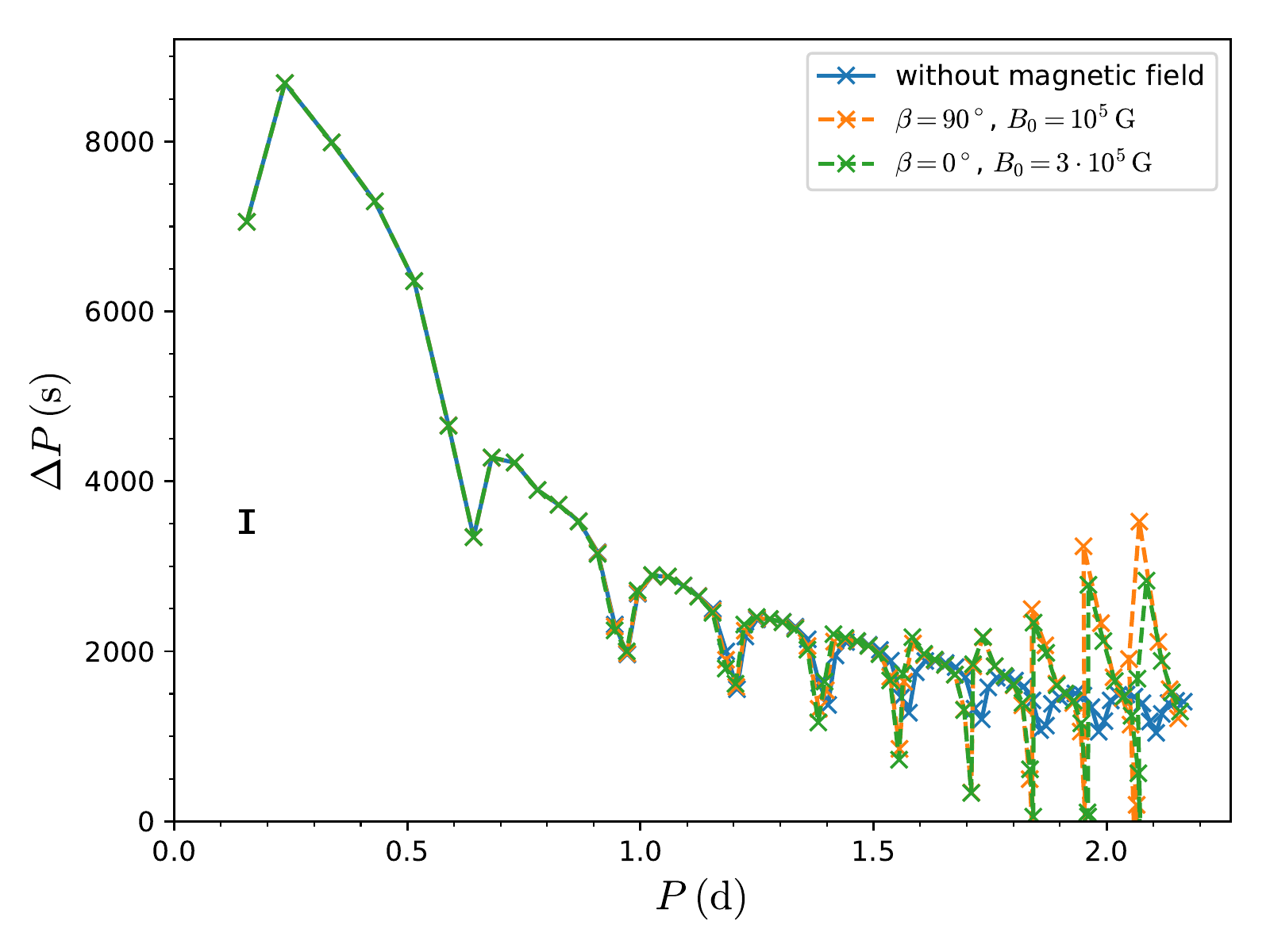}}
    \caption{Same as Fig.~\ref{fig:l1m0} ($m=0$), but with different values of the field strength and the obliquity angle.}
    \label{fig:comp}
\end{figure}

As illustrated in Fig.~\ref{fig:comp}, the enhancement of magnetic frequency shifts obtained when increasing the obliquity angle is similar to the effect of a stronger magnetic field.
However it is not strictly equivalent.
Indeed, the scaling of the frequency shifts induced by the magnetic field with the period for a larger obliquity angle is different from the scaling found for a stronger field.
In theory, it is thus possible to distinguish between the two effects if enough modes are observed (and correctly identified).

\subsection{Prograde modes}
\label{sec:prog}

\begin{figure}
    \resizebox{\hsize}{!}{\includegraphics{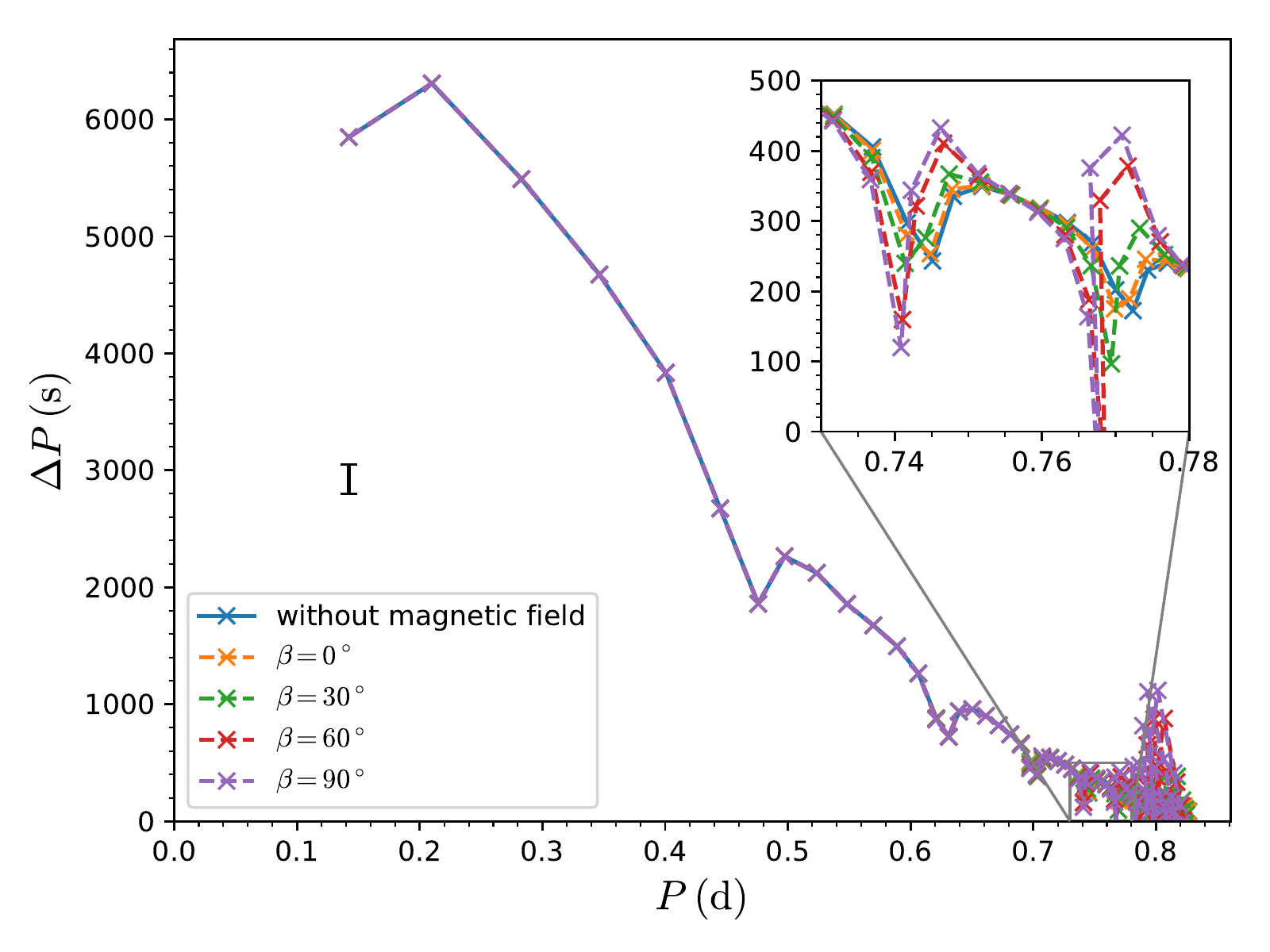}}
    \caption{Same as Fig.~\ref{fig:l1m0}, but for $m=1$.}
    \label{fig:l1m1}
\end{figure}

Figure~\ref{fig:l1m1} represents the period spacings for $m=1$ between modes of consecutive radial orders as a function of the oscillation period.
We observe that oblique magnetic fields have a stronger impact on prograde modes than axisymmetric magnetic fields, similarly to the case of zonal modes.
However, their period spacings go to zero more rapidly for high radial orders than those of zonal modes, which makes magnetic signatures potentially more difficult to detect.
This difficulty might be counterbalanced by the fact that prograde sectoral dipole ($\ell=m=1$) \emph{g} modes are predominantly observed in rapidly rotating intermediate-mass stars \citep[see e.g.][]{VanReeth16, Moravveji16, Papics17, Li19a, Li19b, Li20}.

\subsection{Retrograde modes}
\label{sec:retro}

\begin{figure}
    \resizebox{\hsize}{!}{\includegraphics{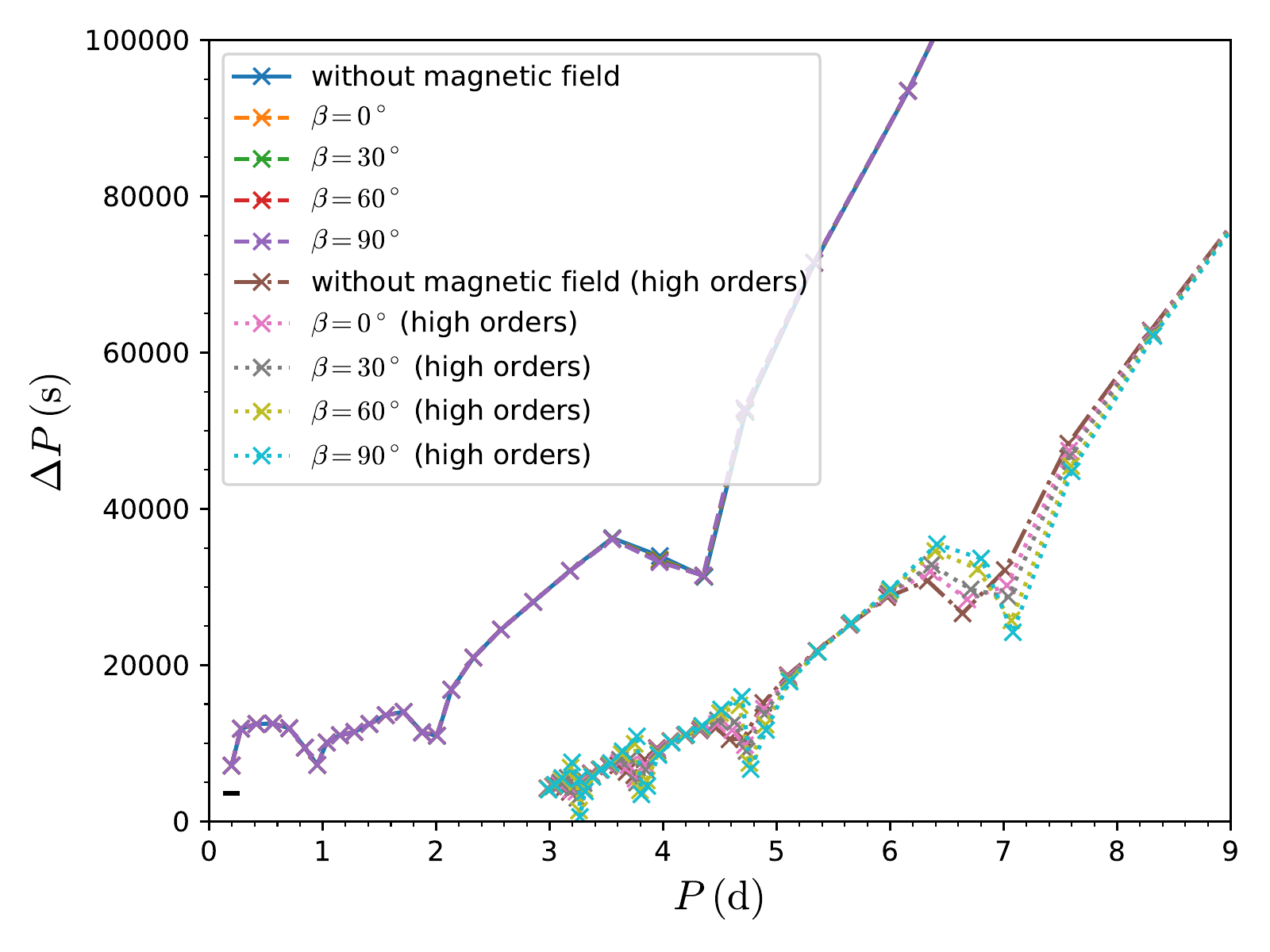}}
    \caption{Same as Fig.~\ref{fig:l1m0}, but for $m=-1$.}
    \label{fig:l1m-1}
\end{figure}

Figure~\ref{fig:l1m-1} represents the period spacings for $m=-1$ between modes of consecutive radial orders as a function of the oscillation period.
Because of the Doppler shift, retrograde modes are split into two branches.
For $m=-1$, this split occurs at $\omega=\Omega$ (in the corotating frame).
The impact of the magnetic field is negligible on the low-order branch, while it is clearly visible on the high-order branch.
Again, oblique fields have stronger signatures than axisymmetric ones.
Most of the modes identified in the spectrum of \hd{} are retrograde modes, but they are all on the low-order branch, and only weak signatures are thus expected.
This is consistent with the absence of a detected observational signature of a magnetic field.

\subsection{Influence of rotation}
\label{sec:rot}

\begin{figure*}
    \resizebox{\hsize}{!}{\includegraphics{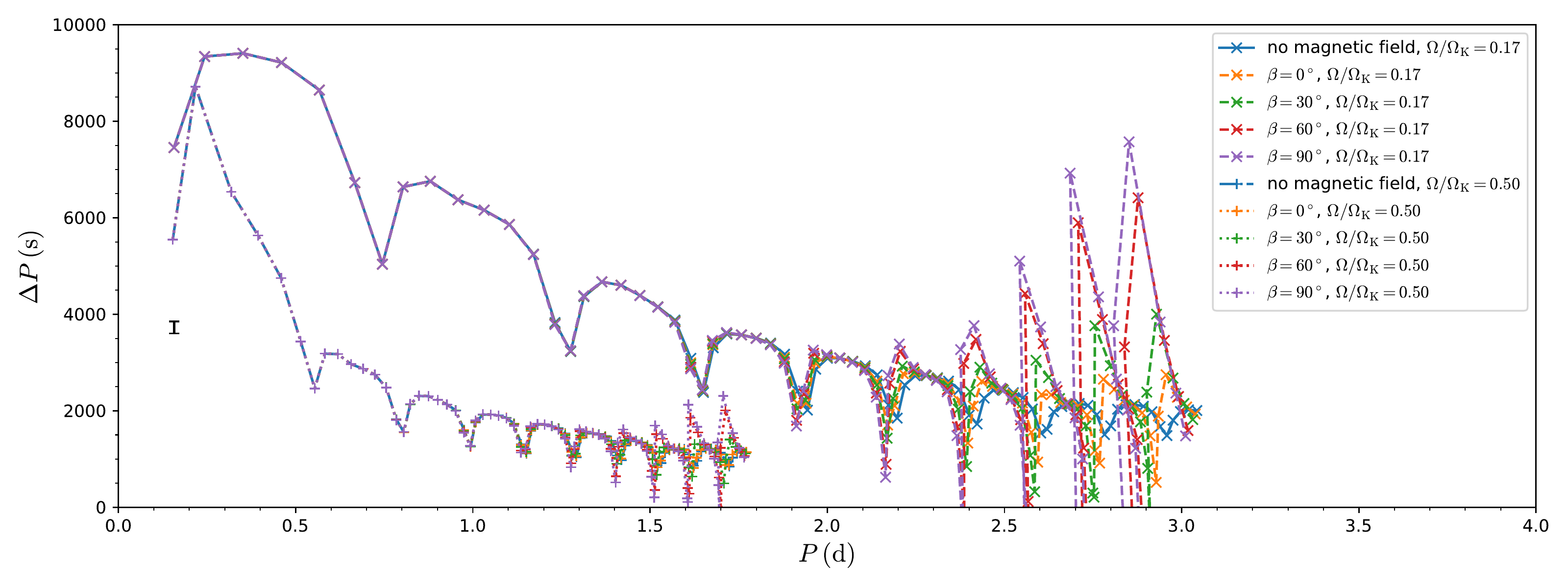}}
    \caption{Same as Fig.~\ref{fig:l1m0} ($m=0$), for two different rotation rates: $\Omega=0.5\Omega_{\star}$ ($\Omega/\Omega_{\rm K}=0.17$) and $\Omega=1.5\Omega_{\star}$ ($\Omega/\Omega_{\rm K}=0.50$), where $\Omega_{\rm K}$ is the Keplerian angular velocity.}
    \label{fig:l1m0_rot}
\end{figure*}

We now investigate the effect of the rotation rate on the frequency shifts induced by oblique fields.
To do so, we computed modes and the associated frequency shifts induced by the magnetic field with 50\% and 150\% of $\Omega_\star$, the measured rotation rate of \hd.
This corresponds to 17\% and 50\% of the Keplerian angular velocity, respectively.
For $\ell=1$ and $m=0$, we obtain the period spacings plotted in Fig.~\ref{fig:l1m0_rot}.
Similarly to the axisymmetric case, the frequency shift induced by the magnetic field for a given mode (at a given radial order) is larger at the lower rotation rate, due to the fact that the mode has a larger period and is thus more sensitive to the magnetic field, according to Eq.~\eqref{eq:freq_ratio}.
In contrast, however, for modes at different rotation rates but close to the same given period, the frequency shift induced by the magnetic field is larger for the larger rotation rate when the obliquity is large.
This comes from the more efficient equatorial trapping of large-period modes for larger rotation rates.

Since oblique fields can generate much larger frequency shifts, they require a smaller amplitude to generate negative period spacing values (visible in Fig.~\ref{fig:l1m0_rot} for the lower rotation rate).
Such negative values mean that the perturbative treatment of the magnetic field is no longer valid.
Indeed, such frequency shifts would lead to avoided crossings between consecutive modes, and the description of such interactions between modes requires a non-perturbative treatment \citep[e.g.]{MorsinkRezania, MathisdeBrye11} of the magnetic field (see also the discussion in Van Beeck et al., submitted).

\section{Discussion and conclusions}
\label{sec:conc}

The analysis presented here shows that the results of Paper~I for an axisymmetric, mixed, internal large-scale magnetic field are still valid for the more realistic case of an oblique field.
Indeed, the magnetic effect on the oscillation frequencies of gravito-inertial modes in the traditional approximation of rotation modifies the period spacing patterns.
The major new result of this study is that the signatures of the presence of a magnetic field in the period spacing patterns are stronger for an oblique field compared to an axisymmetric field, typically up to a factor ten when the field is fully inclined.
This is because the magnetic axis is closer to the equatorial region, where low-frequency modes are trapped.
As a consequence, weaker fields can be detected if they are inclined.
This is true for all mode geometries, including prograde sectoral modes, which are predominantly observed in rapidly rotating stars, and retrograde modes, which are predominantly observed in the particular case of \hd{} \citep{Buysschaert18}.
The change in the patterns due to the strength of the field and to its obliquity can be distinguished from each other if enough modes are identified in the appropriate frequency regime, especially if several period spacing patterns for different mode geometries are observed.
In addition, these signatures are more easily observed if the star rotates slowly.
As a consequence, the search for patterns as those shown here would allow us to indirectly discover internal magnetic fields from observations of gravity modes in hot stars.

The influence of various stellar parameters, such as the mass, the age, and the metallicity, on the expected magnetic signatures is investigated in an extensive parallel paper (Van Beeck et al., submitted).
The main result of this study is that magnetic signatures are detectable in \emph{Kepler} data and should be more easily detectable for stars near the terminal-age main sequence.
A dedicated search for magnetic signatures such as those predicted in this work has not yet been undertaken in \emph{Kepler} data for intermediate-mass \emph{g}-mode pulsators.
Additionally, many hot stars have recently been observed with the TESS mission \citep{Bowman, Pedersen}.
The identification and detailed seismic modelling of pulsating magnetic hot stars will provide observational constraints on the internal properties of their magnetic field.
In addition, including magnetic effects on oscillation frequencies will also provide much improved constraints on stellar interiors for early-type stars.
This would be a new important step into the recent domain of magneto-asteroseismology, but such observational detections have not been made yet.
The new TESS samples of pulsating OB stars are promising in this respect (\citealt{DavidUraz}; Neiner et al., in prep.).

To progress even further on magneto-seismic diagnostics, a non-perturbative treatment of the magnetic field would be needed to be able to consider (i) a magnetic field that is not confined inside the star, and (ii) stronger fields such as those of Ap stars.
Moreover, the formalism presented here could be applied to Rossby modes, which are observed in many intermediate-mass stars \citep{VanReeth16,Saio18, Li19b}, to provide additional constraints on internal magnetic fields of stars.

\begin{acknowledgements}
    The research leading to these results received funding from the European Research Council (ERC) under the European Union's Horizon 2020 research and innovation program (grant agreements No.~647383: SPIRE with PI S.M. and No.~670519: MAMSIE with PI C.A.).
    V.P. and S.M. acknowledge support from the CNES PLATO grant at CEA/DAp.
    The authors thank the anonymous referee for useful comments.
\end{acknowledgements}

\bibliographystyle{aa}
\bibliography{refs}

\begin{thebibliography}{38}
\expandafter\ifx\csname natexlab\endcsname\relax\def\natexlab#1{#1}\fi

\bibitem[{{Aerts} {et~al.}(2019){Aerts}, {Mathis}, \& {Rogers}}]{Aerts19}
{Aerts}, C., {Mathis}, S., \& {Rogers}, T.~M. 2019, \araa, 57, 35

\bibitem[{{Baglin} {et~al.}(2006){Baglin}, {Auvergne}, {Barge}, {Deleuil},
  {Catala}, {Michel}, {Weiss}, \& {COROT Team}}]{Baglin}
{Baglin}, A., {Auvergne}, M., {Barge}, P., {et~al.} 2006, in ESA Special
  Publication, Vol. 1306, The CoRoT Mission Pre-Launch Status - Stellar
  Seismology and Planet Finding, ed. M.~{Fridlund}, A.~{Baglin}, J.~{Lochard},
  \& L.~{Conroy}, 33

\bibitem[{{Borucki} {et~al.}(2010){Borucki}, {Koch}, {Basri}, {Batalha},
  {Brown}, {Caldwell}, {Caldwell}, {Christensen-Dalsgaard}, {Cochran},
  {DeVore}, {Dunham}, {Dupree}, {Gautier}, {Geary}, {Gilliland}, {Gould},
  {Howell}, {Jenkins}, {Kondo}, {Latham}, {Marcy}, {Meibom}, {Kjeldsen},
  {Lissauer}, {Monet}, {Morrison}, {Sasselov}, {Tarter}, {Boss}, {Brownlee},
  {Owen}, {Buzasi}, {Charbonneau}, {Doyle}, {Fortney}, {Ford}, {Holman},
  {Seager}, {Steffen}, {Welsh}, {Rowe}, {Anderson}, {Buchhave}, {Ciardi},
  {Walkowicz}, {Sherry}, {Horch}, {Isaacson}, {Everett}, {Fischer}, {Torres},
  {Johnson}, {Endl}, {MacQueen}, {Bryson}, {Dotson}, {Haas}, {Kolodziejczak},
  {Van Cleve}, {Chandrasekaran}, {Twicken}, {Quintana}, {Clarke}, {Allen},
  {Li}, {Wu}, {Tenenbaum}, {Verner}, {Bruhweiler}, {Barnes}, \&
  {Prsa}}]{Borucki}
{Borucki}, W.~J., {Koch}, D., {Basri}, G., {et~al.} 2010, Science, 327, 977

\bibitem[{{Bouabid} {et~al.}(2013){Bouabid}, {Dupret}, {Salmon},
  {Montalb{\'a}n}, {Miglio}, \& {Noels}}]{Bouabid}
{Bouabid}, M.-P., {Dupret}, M.-A., {Salmon}, S., {et~al.} 2013, \mnras, 429,
  2500

\bibitem[{{Bowman} {et~al.}(2019){Bowman}, {Burssens}, {Pedersen}, {Johnston},
  {Aerts}, {Buysschaert}, {Michielsen}, {Tkachenko}, {Rogers}, {Edelmann},
  {Ratnasingam}, {Sim{\'o}n-D{\'\i}az}, {Castro}, {Moravveji}, {Pope}, {White},
  \& {De Cat}}]{Bowman}
{Bowman}, D.~M., {Burssens}, S., {Pedersen}, M.~G., {et~al.} 2019, Nat.
  Astron., 3, 760

\bibitem[{{Braithwaite}(2008)}]{Braithwaite08}
{Braithwaite}, J. 2008, \mnras, 386, 1947

\bibitem[{{Braithwaite} \& {Spruit}(2004)}]{BraithwaiteSpruit}
{Braithwaite}, J. \& {Spruit}, H.~C. 2004, \nat, 431, 819

\bibitem[{{Buysschaert} {et~al.}(2018){Buysschaert}, {Aerts}, {Bowman},
  {Johnston}, {Van Reeth}, {Pedersen}, {Mathis}, \& {Neiner}}]{Buysschaert18}
{Buysschaert}, B., {Aerts}, C., {Bowman}, D.~M., {et~al.} 2018, \aap, 616, A148

\bibitem[{{Christophe} {et~al.}(2018){Christophe}, {Ballot}, {Ouazzani},
  {Antoci}, \& {Salmon}}]{Christophe}
{Christophe}, S., {Ballot}, J., {Ouazzani}, R.-M., {Antoci}, V., \& {Salmon},
  S.~J.~A.~J. 2018, \aap, 618, A47

\bibitem[{{David-Uraz} {et~al.}(2019){David-Uraz}, {Neiner}, {Sikora},
  {Bowman}, {Petit}, {Chowdhury}, {Handler}, {Pergeorelis}, {Cantiello},
  {Cohen}, {Erba}, {Keszthelyi}, {Khalack}, {Kobzar}, {Kochukhov},
  {Labadie-Bartz}, {Lovekin}, {MacInnis}, {Owocki}, {Pablo}, {Shultz},
  {ud-Doula}, \& {Wade}}]{DavidUraz}
{David-Uraz}, A., {Neiner}, C., {Sikora}, J., {et~al.} 2019, \mnras, 487, 304

\bibitem[{{Donati} \& {Landstreet}(2009)}]{DonatiLandstreet}
{Donati}, J.-F. \& {Landstreet}, J.~D. 2009, \araa, 47, 333

\bibitem[{{Duez} {et~al.}(2010{\natexlab{a}}){Duez}, {Braithwaite}, \&
  {Mathis}}]{DuezBM}
{Duez}, V., {Braithwaite}, J., \& {Mathis}, S. 2010{\natexlab{a}}, \apjl, 724,
  L34

\bibitem[{{Duez} {et~al.}(2010{\natexlab{b}}){Duez}, {Mathis}, \&
  {Turck-Chi{\`e}ze}}]{DuezMTC}
{Duez}, V., {Mathis}, S., \& {Turck-Chi{\`e}ze}, S. 2010{\natexlab{b}}, \mnras,
  402, 271

\bibitem[{{Grunhut} \& {Neiner}(2015)}]{grunhut2015}
{Grunhut}, J.~H. \& {Neiner}, C. 2015, in IAU Symposium, Vol. 305, Polarimetry,
  ed. K.~N. {Nagendra}, S.~{Bagnulo}, R.~{Centeno}, \& M.~{Jes{\'u}s
  Mart{\'\i}nez Gonz{\'a}lez}, 53--60

\bibitem[{{Hough}(1898)}]{Hough}
{Hough}, S.~S. 1898, Philos. T. Roy. Soc. Lon. A, 191, 139

\bibitem[{{Howell} {et~al.}(2014){Howell}, {Sobeck}, {Haas}, {Still},
  {Barclay}, {Mullally}, {Troeltzsch}, {Aigrain}, {Bryson}, {Caldwell},
  {Chaplin}, {Cochran}, {Huber}, {Marcy}, {Miglio}, {Najita}, {Smith},
  {Twicken}, \& {Fortney}}]{Howell}
{Howell}, S.~B., {Sobeck}, C., {Haas}, M., {et~al.} 2014, \pasp, 126, 398

\bibitem[{{Lee} \& {Saio}(1997)}]{LeeSaio97}
{Lee}, U. \& {Saio}, H. 1997, \apj, 491, 839

\bibitem[{{Li} {et~al.}(2019{\natexlab{a}}){Li}, {Bedding}, {Murphy}, {Van
  Reeth}, {Antoci}, \& {Ouazzani}}]{Li19a}
{Li}, G., {Bedding}, T.~R., {Murphy}, S.~J., {et~al.} 2019{\natexlab{a}},
  \mnras, 482, 1757

\bibitem[{{Li} {et~al.}(2019{\natexlab{b}}){Li}, {Van Reeth}, {Bedding},
  {Murphy}, \& {Antoci}}]{Li19b}
{Li}, G., {Van Reeth}, T., {Bedding}, T.~R., {Murphy}, S.~J., \& {Antoci}, V.
  2019{\natexlab{b}}, \mnras, 487, 782

\bibitem[{{Li} {et~al.}(2020){Li}, {Van Reeth}, {Bedding}, {Murphy}, {Antoci},
  {Ouazzani}, \& {Barbara}}]{Li20}
{Li}, G., {Van Reeth}, T., {Bedding}, T.~R., {et~al.} 2020, \mnras, 491, 3586

\bibitem[{{Mathis} \& {de Brye}(2011)}]{MathisdeBrye11}
{Mathis}, S. \& {de Brye}, N. 2011, \aap, 526, A65

\bibitem[{{Moravveji} {et~al.}(2016){Moravveji}, {Townsend}, {Aerts}, \&
  {Mathis}}]{Moravveji16}
{Moravveji}, E., {Townsend}, R.~H.~D., {Aerts}, C., \& {Mathis}, S. 2016, \apj,
  823, 130

\bibitem[{{Morsink} \& {Rezania}(2002)}]{MorsinkRezania}
{Morsink}, S.~M. \& {Rezania}, V. 2002, \apj, 574, 908

\bibitem[{{Neiner} {et~al.}(2015){Neiner}, {Mathis}, {Alecian}, {Emeriau},
  {Grunhut}, {BinaMIcS}, \& {MiMeS Collaborations}}]{Neiner15}
{Neiner}, C., {Mathis}, S., {Alecian}, E., {et~al.} 2015, in IAU Symposium,
  Vol. 305, Polarimetry, ed. K.~N. {Nagendra}, S.~{Bagnulo}, R.~{Centeno}, \&
  M.~{Jes{\'u}s Mart{\'{\i}}nez Gonz{\'a}lez}, 61--66

\bibitem[{{Ouazzani} {et~al.}(2017){Ouazzani}, {Salmon}, {Antoci}, {Bedding},
  {Murphy}, \& {Roxburgh}}]{Ouazzani17}
{Ouazzani}, R.-M., {Salmon}, S.~J.~A.~J., {Antoci}, V., {et~al.} 2017, \mnras,
  465, 2294

\bibitem[{{P{\'a}pics} {et~al.}(2012){P{\'a}pics}, {Briquet}, {Baglin},
  {Poretti}, {Aerts}, {Degroote}, {Tkachenko}, {Morel}, {Zima}, {Niemczura},
  {Rainer}, {Hareter}, {Baudin}, {Catala}, {Michel}, {Samadi}, \&
  {Auvergne}}]{Papics12}
{P{\'a}pics}, P.~I., {Briquet}, M., {Baglin}, A., {et~al.} 2012, \aap, 542, A55

\bibitem[{{P{\'a}pics} {et~al.}(2017){P{\'a}pics}, {Tkachenko}, {Van Reeth},
  {Aerts}, {Moravveji}, {Van de Sande}, {De Smedt}, {Bloemen}, {Southworth},
  {Debosscher}, {Niemczura}, \& {Gameiro}}]{Papics17}
{P{\'a}pics}, P.~I., {Tkachenko}, A., {Van Reeth}, T., {et~al.} 2017, \aap,
  598, A74

\bibitem[{{Pedersen} {et~al.}(2019){Pedersen}, {Chowdhury}, {Johnston},
  {Bowman}, {Aerts}, {Handler}, {De Cat}, {Neiner}, {David-Uraz}, {Buzasi},
  {Tkachenko}, {Sim{\'o}n-D{\'\i}az}, {Moravveji}, {Sikora}, {Mirouh},
  {Lovekin}, {Cantiello}, {Daszy{\'n}ska-Daszkiewicz}, {Pigulski},
  {Vanderspek}, \& {Ricker}}]{Pedersen}
{Pedersen}, M.~G., {Chowdhury}, S., {Johnston}, C., {et~al.} 2019, \apjl, 872,
  L9

\bibitem[{{Prat} {et~al.}(2019){Prat}, {Mathis}, {Buysschaert}, {Van Beeck},
  {Bowman}, {Aerts}, \& {Neiner}}]{paper1}
{Prat}, V., {Mathis}, S., {Buysschaert}, B., {et~al.} 2019, \aap, 627, A64,
  paper I

\bibitem[{{Ricker} {et~al.}(2015){Ricker}, {Winn}, {Vanderspek}, {Latham},
  {Bakos}, {Bean}, {Berta-Thompson}, {Brown}, {Buchhave}, {Butler}, {Butler},
  {Chaplin}, {Charbonneau}, {Christensen-Dalsgaard}, {Clampin}, {Deming},
  {Doty}, {De Lee}, {Dressing}, {Dunham}, {Endl}, {Fressin}, {Ge}, {Henning},
  {Holman}, {Howard}, {Ida}, {Jenkins}, {Jernigan}, {Johnson}, {Kaltenegger},
  {Kawai}, {Kjeldsen}, {Laughlin}, {Levine}, {Lin}, {Lissauer}, {MacQueen},
  {Marcy}, {McCullough}, {Morton}, {Narita}, {Paegert}, {Palle}, {Pepe},
  {Pepper}, {Quirrenbach}, {Rinehart}, {Sasselov}, {Sato}, {Seager},
  {Sozzetti}, {Stassun}, {Sullivan}, {Szentgyorgyi}, {Torres}, {Udry}, \&
  {Villasenor}}]{Ricker}
{Ricker}, G.~R., {Winn}, J.~N., {Vanderspek}, R., {et~al.} 2015, J. Astron.
  Telesc. Instrum. Syst., 1, 014003

\bibitem[{{Saio} {et~al.}(2018){Saio}, {Kurtz}, {Murphy}, {Antoci}, \&
  {Lee}}]{Saio18}
{Saio}, H., {Kurtz}, D.~W., {Murphy}, S.~J., {Antoci}, V.~L., \& {Lee}, U.
  2018, \mnras, 474, 2774

\bibitem[{{Shultz} {et~al.}(2019){Shultz}, {Wade}, {Rivinius}, {Alecian},
  {Neiner}, {Petit}, {Owocki}, {ud-Doula}, {Kochukhov}, {Bohlender},
  {Keszthelyi}, {MiMeS Collaboration}, \& {BinaMIcS
  Collaboration}}]{shultz2019}
{Shultz}, M.~E., {Wade}, G.~A., {Rivinius}, T., {et~al.} 2019, \mnras, 490, 274

\bibitem[{{Tayler}(1980)}]{Tayler80}
{Tayler}, R.~J. 1980, \mnras, 191, 151

\bibitem[{{Townsend}(2003)}]{Townsend03}
{Townsend}, R.~H.~D. 2003, \mnras, 340, 1020

\bibitem[{{Van Reeth} {et~al.}(2016){Van Reeth}, {Tkachenko}, \&
  {Aerts}}]{VanReeth16}
{Van Reeth}, T., {Tkachenko}, A., \& {Aerts}, C. 2016, \aap, 593, A120

\bibitem[{{Van Reeth} {et~al.}(2015{\natexlab{a}}){Van Reeth}, {Tkachenko},
  {Aerts}, {P{\'a}pics}, {Degroote}, {Debosscher}, {Zwintz}, {Bloemen}, {De
  Smedt}, {Hrudkova}, {Raskin}, \& {Van Winckel}}]{VanReeth15a}
{Van Reeth}, T., {Tkachenko}, A., {Aerts}, C., {et~al.} 2015{\natexlab{a}},
  \aap, 574, A17

\bibitem[{{Van Reeth} {et~al.}(2015{\natexlab{b}}){Van Reeth}, {Tkachenko},
  {Aerts}, {P{\'a}pics}, {Triana}, {Zwintz}, {Degroote}, {Debosscher},
  {Bloemen}, {Schmid}, {De Smedt}, {Fremat}, {Fuentes}, {Homan}, {Hrudkova},
  {Karjalainen}, {Lombaert}, {Nemeth}, {{\O}stensen}, {Van De Steene}, {Vos},
  {Raskin}, \& {Van Winckel}}]{VanReeth15b}
{Van Reeth}, T., {Tkachenko}, A., {Aerts}, C., {et~al.} 2015{\natexlab{b}},
  \apjs, 218, 27

\bibitem[{{Wade} {et~al.}(2016){Wade}, {Neiner}, {Alecian}, {Grunhut}, {Petit},
  {Batz}, {Bohlender}, {Cohen}, {Henrichs}, {Kochukhov}, {Landstreet},
  {Manset}, {Martins}, {Mathis}, {Oksala}, {Owocki}, {Rivinius}, {Shultz},
  {Sundqvist}, {Townsend}, {ud-Doula}, {Bouret}, {Braithwaite}, {Briquet},
  {Carciofi}, {David-Uraz}, {Folsom}, {Fullerton}, {Leroy}, {Marcolino},
  {Moffat}, {Naz{\'e}}, {Louis}, {Auri{\`e}re}, {Bagnulo}, {Bailey},
  {Barb{\'a}}, {Blaz{\`e}re}, {B{\"o}hm}, {Catala}, {Donati}, {Ferrario},
  {Harrington}, {Howarth}, {Ignace}, {Kaper}, {L{\"u}ftinger}, {Prinja},
  {Vink}, {Weiss}, \& {Yakunin}}]{Wade}
{Wade}, G.~A., {Neiner}, C., {Alecian}, E., {et~al.} 2016, \mnras, 456, 2

\end{thebibliography}

\appendix
\onecolumn

\section{Non-zero-average terms of the Lorentz work}
\label{sec:terms}

In this section, we list all non-zero-average terms of the work of the Lorentz force induced by a totally oblique dipolar field $\delta\vec F_{\rm L}^{\rm eq}\cdot\vec\xi^*$ (for convenience, we take here the product with $\mu_0$).
Interestingly, all these terms involve either only poloidal components of the magnetic field, or only the toroidal component.
Therefore, we group them accordingly.
All terms have a purely radial part multiplied by a purely latitudinal part, and the prime symbol ($'$) denotes a total derivative, either radial or latitudinal depending on the context.

\subsection{Poloidal terms}

Noting $A = [(rb_\theta)'+b_{\rm r}]$, the terms involving poloidal components are
\begin{align*}
    &\frac{(r\xi_{\rm r}b_\theta)'A\xi_{\rm r}^*}{r^2}H_{\rm r}^2\sin^2\varphi
    +\frac{\xi_{\rm h}b_\theta A\xi_{\rm r}^*}{r^2}H_{\rm r}H_\theta'\sin^2\varphi
    +\frac{\xi_{\rm h}b_\theta A\xi_{\rm r}^*}{r^2}H_{\rm r}H_\theta\frac{\cos\theta}{\sin\theta}\cos^2\varphi
    -m\frac{\xi_{\rm h}b_\theta A\xi_{\rm r}^*}{r^2}H_{\rm r}H_\varphi\frac{\cos^2\theta}{\sin\theta}\cos^2\varphi  \\
    &+\frac{(r\xi_{\rm r}b_\theta)'A\xi_{\rm r}^*}{r^2}H_{\rm r}^2\cos^2\theta\cos^2\varphi
    +\frac{(r\xi_{\rm h}b_{\rm r})'A\xi_{\rm r}^*}{r^2}H_{\rm r}H_\theta\sin\theta\cos\theta\cos^2\varphi
    +\frac{\xi_{\rm r}b_\theta A\xi_{\rm h}^*}{r^2}H_\theta\frac{\cos\theta}{\sin\theta}(H_{\rm r}\sin\theta\cos\theta)'\cos^2\varphi    \\
    &+\frac{|\xi_{\rm h}|^2b_{\rm r}A}{r^2}\frac{(H_\theta\sin^2\theta)'\cos\theta}{\sin\theta}H_\theta\cos^2\varphi
    -m\frac{|\xi_{\rm h}|^2b_{\rm r}A}{r^2}H_\theta H_\varphi\cos\theta\cos^2\varphi
    -\frac{\xi_{\rm r}b_\theta A\xi_{\rm h}^*}{r^2}H_{\rm r}H_\theta\frac{\cos\theta}{\sin\theta}\cos^2\varphi
    +\frac{|\xi_{\rm h}|^2b_{\rm r}A}{r^2}H_\varphi^2\sin^2\varphi  \\
    &+m\frac{\xi_{\rm r}b_\theta A\xi_{\rm h}^*}{r^2}\frac{H_{\rm r}H_\varphi}{\sin\theta}\sin^2\varphi
    +\frac{|\xi_{\rm r}|^2b_\theta^2}{r^2}\frac{(H_{\rm r}\sin\theta\cos\theta)'}{\sin^2\theta}H_{\rm r}\sin^2\varphi
    +\frac{\xi_{\rm h}b_{\rm r}b_\theta\xi_{\rm r}^*}{r^2}\frac{(H_\theta\sin^2\theta)'}{\sin^2\theta}H_{\rm r}\sin^2\varphi
    -2m\frac{\xi_{\rm h}b_{\rm r}b_\theta\xi_{\rm r}^*}{r^2}\frac{H_\varphi H_{\rm r}}{\sin\theta}\sin^2\varphi \\
    &-(1+m^2)\frac{|\xi_{\rm r}|^2b_\theta^2}{r^2}\frac{H_{\rm r}^2}{\sin^2\theta}\sin^2\varphi
    +\frac{(r\xi_{\rm r}b_\theta)''b_\theta\xi_{\rm r}^*}{r}H_{\rm r}^2\sin^2\varphi
    +\frac{(\xi_{\rm h}b_\theta)'b_\theta\xi_{\rm r}^*}{r}H_\theta'H_{\rm r}\sin^2\varphi
    +\frac{(\xi_{\rm h}b_\theta)'b_\theta\xi_{\rm r}^*}{r}H_\theta H_{\rm r}\frac{\cos\theta}{\sin\theta}\cos^2\varphi  \\
    &-m\frac{(\xi_{\rm h}b_\theta)'b_\theta\xi_{\rm }^*}{r}H_\varphi H_{\rm r}\frac{\cos^2\theta}{\sin\theta}\cos^2\varphi
    +\frac{(r\xi_{\rm r}b_\theta)''b_\theta\xi_{\rm r}^*}{r}H_{\rm r}^2\cos^2\theta\cos^2\varphi
    +\frac{(r\xi_{\rm h}b_{\rm r})''b_\theta\xi_{\rm r}^*}{r}H_\theta H_{\rm r}\sin\theta\cos\theta\cos^2\varphi   \\
    &+\frac{|\xi_{\rm r}|^2b_\theta^2}{r^2}\left[\frac{(H_{\rm r}\sin\theta\cos\theta)'}{\sin\theta}\right]'H_{\rm r}\cos\theta\cos^2\varphi
    +\frac{\xi_{\rm h}b_{\rm r}b_\theta\xi_{\rm r}^*}{r^2}\left[\frac{(H_\theta\sin^2\theta)'}{\sin\theta}\right]'H_{\rm r}\cos\theta\cos^2\varphi
    -m\frac{\xi_{\rm h}b_{\rm r}b_\theta\xi_{\rm r}^*}{r^2}H_\varphi'H_{\rm r}\cos\theta\cos^2\varphi   \\
    &-\frac{\xi_{\rm r}|^2b_\theta^2}{r^2}\left(\frac{H_{\rm r}}{\sin\theta}\right)'H_{\rm r}\cos\theta\cos^2\varphi
    +\frac{(\xi_{\rm h}b_\theta)'b_{\rm r}\xi_{\rm h}^*}{r}H_\theta^2\cos^2\varphi
    -m\frac{(\xi_{\rm h}b_\theta)'b_{\rm r}\xi_{\rm h}^*}{r}H_\varphi H_\theta\cos\theta\cos^2\varphi
    +\frac{(r\xi_{\rm r}b_\theta)''b_{\rm r}\xi_{\rm h}^*}{r}H_{\rm r}H_\theta\sin\theta\cos\theta\cos^2\varphi \\
    &+\frac{(r\xi_{\rm h}b_{\rm r})''b_{\rm r}\xi_{\rm h}^*}{r}H_\theta^2\sin^2\theta\cos^2\varphi
    +\frac{\xi_{\rm r}b_\theta b_{\rm r}\xi_{\rm h}^*}{r^2}\left[\frac{(H_{\rm r}\sin\theta\cos\theta)'}{\sin\theta}\right]'H_\theta\sin\theta\cos^2\varphi
    +\frac{|\xi_{\rm h}|^2b_{\rm r}^2}{r^2}\left[\frac{(H_\theta\sin^2\theta)'}{\sin\theta}\right]'H_\theta\sin\theta\cos^2\varphi  \\
    &-m\frac{|\xi_{\rm h}|^2b_{\rm r}^2}{r^2}H_\varphi'H_\theta\sin\theta\cos^2\varphi
    -\frac{\xi_{\rm r}b_\theta b_{\rm r}\xi_{\rm h}^*}{r^2}\left(\frac{H_{\rm r}}{\sin\theta}\right)'H_\theta\sin\theta\cos^2\varphi
    +\frac{(r\xi_{\rm r}b_\theta)'b_\theta\xi_{\rm h}^*}{r^2}\frac{(H_{\rm r}\sin\theta)'H_\theta}{\sin\theta}\sin^2\varphi
    +\frac{|\xi_{\rm h}|^2b_\theta^2}{r^2}\frac{(H_\theta'\sin\theta)'H_\theta}{\sin\theta}\sin^2\varphi    \\
    &-(1+m^2)\frac{|\xi_{\rm h}|^2b_\theta^2}{r^2}\frac{H_\theta^2}{\sin^2\theta}\sin^2\varphi
    +2m\frac{|\xi_{\rm h}|^2b_\theta^2}{r^2}\frac{H_\varphi H_\theta\cos\theta}{\sin^2\theta}\sin^2\varphi
    -\frac{(r\xi_{\rm r}b_\theta)'b_\theta\xi_{\rm h}^*}{r^2}H_{\rm r}H_\theta\frac{\cos\theta}{\sin\theta}\sin^2\varphi
    -\frac{(r\xi_{\rm h}b_{\rm r})'b_\theta\xi_{\rm h}^*}{r^2}H_\theta^2\sin^2\varphi   \\
    &-\frac{(r\xi_{\rm h}b_{\rm r})'b_\theta\xi_{\rm h}^*}{r^2}\frac{(H_\varphi\sin^2\theta)'H_\varphi\cos\theta}{\sin\theta}\cos^2\varphi
    +\frac{|\xi_{\rm h}|^2b_\theta^2}{r^2}\frac{[(H_\varphi\cos\theta)'\sin\theta]'H_\varphi\cos\theta}{\sin\theta}\cos^2\varphi
    +2m\frac{|\xi_{\rm h}|^2b_\theta^2}{r^2}\frac{H_\theta H_\varphi\cos\theta}{\sin^2\theta}\cos^2\varphi  \\
    &-(1+m^2)\frac{|\xi_{\rm h}|^2b_\theta^2}{r^2}\frac{H_\varphi^2\cos^2\theta}{\sin^2\theta}\cos^2\varphi
    +m\frac{(r\xi_{\rm r}b_\theta)'b_\theta\xi_{\rm h}^*}{r^2}H_{\rm r}H_\varphi\frac{\cos^2\theta}{\sin\theta}\cos^2\varphi
    +m\frac{(r\xi_{\rm h}b_{\rm r})'b_\theta\xi_{\rm h}^*}{r^2}H_\theta H_\varphi\cos\theta\cos^2\varphi  \\
    &+m\frac{\xi_{\rm r}b_\theta b_{\rm r}\xi_{\rm h}^*}{r^2}\frac{(H_{\rm r}\sin\theta\cos\theta)'H_\varphi}{\sin\theta}\cos^2\varphi
    +m\frac{|\xi_{\rm h}|^2b_{\rm r}^2}{r^2}\frac{(H_\theta\sin^2\theta)'H_\varphi}{\sin\theta}\sin^2\varphi
    -(1+m^2)\frac{|\xi_{\rm h}|^2b_{\rm r}^2}{r^2}H_\varphi^2\cos^2\varphi
    -2m\frac{\xi_{\rm r}b_\theta b_{\rm r}\xi_{\rm h}^*}{r^2}\frac{H_{\rm r}H_\varphi}{\sin\theta}\cos^2\varphi \\
    &+\frac{(r\xi_{\rm h}b_{\rm r})''b_{\rm r}\xi_{\rm h}^*}{r}H_\varphi^2\sin^2\theta\cos^2\varphi
    -\frac{(\xi_{\rm h}b_\theta)'b_{\rm r}\xi_{\rm h}^*}{r}(H_\varphi\cos\theta)'H_\varphi\sin\theta\cos^2\varphi.
\end{align*}

\subsection{Toroidal terms}
\label{sec:tor}

The terms that involve the toroidal component are
\begin{align*}
    &\frac{(r\xi_{\rm r}b_\varphi)'(rb_\varphi)'\xi_{\rm r}^*}{r^2}H_{\rm r}^2\cos^2\theta\cos^2\varphi
    +\frac{\xi_{\rm h}b_\varphi(rb_\varphi)'\xi_{\rm r}^*}{r^2}H_{\rm r}(H_\theta\cos\theta)'\cos\theta\cos^2\varphi
    +\frac{\xi_{\rm h}b_\varphi(rb_\varphi)'\xi_{\rm r}^*}{r^2}H_{\rm r}H_\theta\frac{\cos\theta}{\sin\theta}\sin^2\varphi   \\
    &-m\frac{\xi_{\rm h}b_\varphi(rb_\varphi)'\xi_{\rm }^*}{r^2}\frac{H_{\rm r}H_\varphi}{\sin\theta}\sin^2\varphi
    +\frac{(r\xi_{\rm r}b_\varphi)'(rb_\varphi)'\xi_{\rm r}^*}{r^2}H_{\rm r}^2\sin^2\varphi
    +\frac{\xi_{\rm r}b_\varphi(rb_\varphi)'\xi_{\rm h}^*}{r^2}H_\theta\frac{(H_{\rm r}\sin\theta)'}{\sin\theta}\sin^2\varphi
    -\frac{\xi_{\rm r}b_\varphi(rb_\varphi)'\xi_{\rm h}^*}{r^2}H_{\rm r}H_\theta\frac{\cos\theta}{\sin\theta}\sin^2\varphi  \\
    &-2\frac{(r\xi_{\rm r}b_\varphi)'b_\varphi\xi_{\rm h}^*}{r^2}H_{\rm r}H_\theta\sin\theta\cos\theta\cos^2\varphi
    -2\frac{|\xi_{\rm h}|^2b_\varphi^2}{r^2}H_\theta(H_\theta\cos\theta)'\sin\theta\cos^2\varphi
    -2m\frac{|\xi_{\rm h}|^2b_\varphi^2}{r^2}H_\theta H_\varphi\cos\theta\cos^2\varphi
    +2\frac{|\xi_{\rm h}|^2b_\varphi^2}{r^2}H_\varphi^2\cos^2\varphi    \\
    &m\frac{\xi_{\rm r}b_\varphi(rb_\varphi)'\xi_{\rm h}^*}{r^2}H_{\rm r}H_\varphi\frac{\cos^2\theta}{\sin\theta}\cos^2\varphi
    +\frac{|\xi_{\rm r}|^2b_\varphi^2}{r^2}\frac{(H_{\rm r}\sin\theta)'}{\sin^2\theta}H_{\rm r}\cos\theta\cos^2\varphi
    -(1+m^2)\frac{|\xi_{\rm r}|^2b_\varphi^2}{r^2}H_{\rm r}^2\frac{\cos^2\theta}{\sin^2\theta}\cos^2\varphi \\
    &+\frac{(r\xi_{\rm r}b_\varphi)''b_\varphi\xi_{\rm r}^*}{r}H_{\rm r}^2\cos^2\theta\cos^2\varphi
    +\frac{(\xi_{\rm h}b_\varphi)'b_\varphi\xi_{\rm r}^*}{r}(H_\theta\cos\theta)'H_{\rm r}\cos\theta\cos^2\varphi
    +\frac{(\xi_{\rm h}b_\varphi)'b_\varphi\xi_{\rm r}^*}{r}H_\theta H_{\rm r}\frac{\cos\theta}{\sin\theta}\sin^2\varphi \\
    &-m\frac{(\xi_{\rm h}b_\varphi)'b_\varphi\xi_{\rm r}^*}{r}\frac{H_\varphi H_{\rm r}}{\sin\theta}\sin^2\varphi
    +\frac{(r\xi_{\rm r}b_\varphi)''b_\varphi\xi_{\rm r}^*}{r}H_{\rm r}^2\sin^2\varphi^2
    +\frac{|\xi_{\rm r}|^2b_\varphi^2}{r^2}\left[\frac{(H_{\rm r}\sin\theta)'}{\sin\theta}\right]'H_{\rm r}\sin^2\varphi
    -\frac{|\xi_{\rm r}|^2b_\varphi^2}{r^2}\left(H_{\rm r}\frac{\cos\theta}{\sin\theta}\right)'H_{\rm r}\sin^2\varphi   \\
    &+\frac{(r\xi_{\rm r}b_\varphi)'b_\varphi\xi_{\rm h}^*}{r^2}\frac{(H_{\rm r}\sin\theta\cos\theta)'H_\theta\cos\theta}{\sin\theta}\cos^2\varphi
    +\frac{|\xi_{\rm h}|^2b_\varphi^2}{r^2}\frac{[(H_\theta\cos\theta)'\sin\theta]'H_\theta\cos\theta}{\sin\theta}\cos^2\varphi
    -(1+m^2)\frac{|\xi_{\rm h}|^2b_\varphi^2}{r^2}H_\theta^2\frac{\cos^2\theta}{\sin^2\theta}\cos^2\varphi  \\
    &+2m\frac{|\xi_{\rm h}|^2b_\varphi^2}{r^2}\frac{H_\varphi H_\theta\cos\theta}{\sin^2\theta}\cos^2\varphi
    -\frac{(r\xi_{\rm r}b_\varphi)'b_\varphi\xi_{\rm h}^*}{r^2}\frac{H_{\rm r}H_\theta\cos\theta}{\sin\theta}\cos^2\varphi
    +\frac{|\xi_{\rm h}|^2b_\varphi^2}{r^2}\frac{(H_\varphi'\sin\theta)'H_\varphi}{\sin\theta}\sin^2\varphi
    +2m\frac{|\xi_{\rm h}|^2b_\varphi^2}{r^2}\frac{H_\theta H_\varphi\cos\theta}{\sin^2\theta}\sin^2\varphi \\
    &-(1+m^2)\frac{|\xi_{\rm h}|^2b_\varphi^2}{r^2}\frac{H_\varphi^2}{\sin^2\theta}\sin^2\varphi
    +m\frac{(r\xi_{\rm r}b_\varphi)'b_\varphi\xi_{\rm h}^*}{r^2}\frac{H_{\rm r}H_\varphi}{\sin\theta}\sin^2\varphi.
\end{align*}

\end{document}